\documentclass[12pt,preprint]{aastex}
\usepackage{float,epsfig}
\newbox\grsign \setbox\grsign=\hbox{$>$} 
\newdimen\grdimen \grdimen=\ht\grsign
\newbox\laxbox \newbox\gaxbox
\setbox\gaxbox=\hbox{\raise.5ex\hbox{$>$}\llap
     {\lower.5ex\hbox{$\sim$}}}\ht1=\grdimen\dp1=0pt
\setbox\laxbox=\hbox{\raise.5ex\hbox{$<$}\llap
     {\lower.5ex\hbox{$\sim$}}}\ht2=\grdimen\dp2=0pt

\def\ltsima{$\; \buildrel < \over \sim \;$}
\def\simlt{\lower.5ex\hbox{\ltsima}} 

\newcommand{\cha}{{\sl Chandra}}

\newcommand{\Fermi}{{\sl Fermi}}

\newcommand{\eV}{${\rm e\!V}$}
\newcommand{\keV}{${\rm ke\!V}$}
\newcommand{\MeV}{${\rm Me\!V}$}
\newcommand{\GeV}{${\rm Ge\!V}$}
\newcommand{\TeV}{${\rm Te\!V}$}

\begin{document}

\title{\sl Chandra, Keck and VLA Observations of the Crab Nebula during the 2011-April Gamma-ray Flare}

\author{
Martin~C.~Weisskopf\altaffilmark{1},
Allyn~F.~Tennant\altaffilmark{1},
Jonathan Arons\altaffilmark{2},
Roger Blandford\altaffilmark{3},\\
Rolf Buehler\altaffilmark{4},
Patrizia Caraveo\altaffilmark{5},
C.~C.~Teddy Cheung\altaffilmark{6},
Enrico Costa\altaffilmark{7},
Andrea de Luca\altaffilmark{5},
Carlo Ferrigno\altaffilmark{8},
Hai Fu\altaffilmark{9},
Stefan Funk\altaffilmark{3},
Moritz Habermehl\altaffilmark{10},
Dieter Horns\altaffilmark{10},\\
Justin~D.~Linford\altaffilmark{11},
Andrei Lobanov\altaffilmark{12},
Claire Max\altaffilmark{13},
Roberto Mignani\altaffilmark{14},\\
Stephen~L.~O'Dell\altaffilmark{1},
Roger W. Romani\altaffilmark{3},
Edoardo Striani\altaffilmark{7},
Marco Tavani\altaffilmark{7},\\
Gregory~B.~Taylor\altaffilmark{11},
Yasunobu Uchiyama\altaffilmark{3},
Yajie Yuan\altaffilmark{3}
}

\altaffiltext{1}
{NASA Marshall Space Flight Center, Astrophysics Office (ZP12), Huntsville, AL 35812, USA}
\altaffiltext{2}
{Astronomy Department and Theoretical Astrophysics Center, University of California, Berkeley, 601 Campbell Hall, Berkeley, CA 94720, USA}
\altaffiltext{3}
{W. W. Hansen Experimental Physics Laboratory, Kavli Institute for Particle Astrophysics and Cosmology, Department of Physics and SLAC National Accelerator Laboratory, Stanford University, Stanford, CA 94305, USA}
\altaffiltext{4}
{DESY, Platanenallee 6, 15738 Zeuthen, Germany}
\altaffiltext{5}
{INAF-IASF Milano, via E. Bassini 15, 20133 Milano, Italy \& INFN Pavia, via A. Bassi 6, 27100 Pavia, Italy}
\altaffiltext{6}
{National Research Council Research Associate, National Academy of Sciences, Washington, DC 20001, resident at Naval Research Laboratory, Washington, DC 20375, USA}
\altaffiltext{7}
{INFN Roma Tor Vergata, via della Ricerca Scientifica 1, 00133 Roma, Italy  }
\altaffiltext{8}
{ISDC, Data Center for Astrophysics of the University of Geneva, chemin d'Écogia 16, 1290, Versoix, Switzerland}
\altaffiltext{9}
{Department of Physics \& Astronomy, University of California, Irvine, CA 92697, USA}
\altaffiltext{10}
{Institut f\"ur Experimentalphysik, Universit\"at Hamburg, Luruper
Chaussee 149, D-22761 Hamburg, Germany}
\altaffiltext{11}
{Department of Physics and Astronomy, University of New Mexico, MSC07 4220, Albuquerque, NM 87131-0001, USA.}
\altaffiltext{12}
{Max-Planck-Institut f{\"u}r Radioastronomie, Auf dem Hügel 69, 53121, Bonn, Germany}
\altaffiltext{13}
{Department of Astronomy \& Astrophysics, University of California, Santa Cruz, CA 95064, USA}
\altaffiltext{14}
{Mullard Space Science Laboratory, University College London, Holmbury St. Mary Dorking, Surrey RH5 6NT, England \& Kepler Institute of Astronomy, University of Zielona G\'ora, Lubuska 2, 65-265, Zielona G\'ora, Poland}

\begin{abstract}
We present results from our analysis of \cha~ X-ray Observatory, W. M. Keck Observatory, and Karl G.~Jansky Very Large Array (VLA) images of the Crab Nebula that were contemporaneous with the $\gamma$-ray flare of 2011 April.
Despite hints in the X-ray data, we find no evidence for statistically significant variations that pinpoint the specific location of the flares within the Nebula. 
The Keck observations extend this conclusion to the ``inner knot", i.e., the feature within an arcsecond of the pulsar.
The VLA observations support this conclusion. 
We also discuss theoretical implications of the $\gamma$-ray flares and suggest that the most dramatic 
$\gamma$-ray flares are due to radiation-reaction-limited synchrotron emission associated with sudden, dissipative changes in the current system sustained by the central pulsar. 
\end{abstract}

\section{Introduction} \label{s:intros}

The Crab Nebula, the relic of a stellar explosion recorded by Chinese astronomers in 1054, has a special place in the history of astronomy. 
It is our most frequently observed laboratory for high-energy astrophysics. 
Located at a distance of $\approx2$ kpc, the system is energized by a pulsar of spindown luminosity $L_{plsr}\approx5\times 10^{38}$ erg/s and current spin period P $\approx34$ ms. 
The history and general properties of the system are nicely summarized in the review by Hester (2008). 
Optical and X-ray images (Hester et al.\ 1995; Weisskopf et al.\ 2000; Hester et al.\ 2002) of the inner nebula show features such as an inner ring, toroidal structure, knots, and two opposing jets originating from the pulsar –-- these latter presumably aligned with its rotation axis and proper motion vector (Caraveo \& Mignani 1999; Ng \& Romani 2007; Kaplan et al.\ 2008 and references therein). 
The ``inner-ring", prominent in X-rays, is commonly accepted as being the termination shock produced by the relativistic wind of particles accelerated by the pulsar. 
Many of the optical and X-ray features brighten and fade and/or move over weeks or months (e.g., Hester et al.\ 1995; Hester et al.\ 2002).

The quiescent or average spectral energy distribution (SED) of the Crab Nebula has a characteristic two-humped form (see, e.g., Figure~\ref{f:sed}; Atoyan \& Aharonian 1996; Bucciantini, Arons, \& Amato 2011 and references therein).
The synchrotron spectrum extends from $\approx30$~MHz to $\approx 1.2\times 10^{22}$~Hz (500~\MeV). 
Most of the power is radiated  by $\approx$~\TeV\ electrons in the near UV $\approx10$~\eV\ with an associated luminosity of $\approx 1.3 \times 10^{38}$ {\rm ergs/s} (Hester 2008). 
This roughly matches the loss of rotational energy by the pulsar, which releases its energy electromagnetically,  generating a current $\approx200$~TA and inducing an electro-motive force (EMF) $\approx50$~PV. 
However, the nebula is currently varying on a few-year timescale (Wilson-Hodge et al.\ 2011).
At higher energies, Compton scattering has a luminosity of $\approx10^{36}$~{\rm erg/s}, peaking around 60~\GeV\ (Albert et al.\ 2008) and measured up to $\approx80$~\TeV\ (e.g., Aharonian et al.\ 2004; Abdo et al.\ 2010). 

Since 2007, the AGILE and \Fermi~ satellites have detected several $\gamma$-ray flares from the Crab Nebula (Tavani et al.\ 2011; Abdo et al.\ 2011; Striani et al.\ 2011a; Buehler et al.\ 2012) in the $0.1-1$~\GeV\ range.  
The most dramatic flares exhibit variability on timescales as short as a few hours, although it is unclear whether they are distinct events or just the largest variations from a stationary power spectrum of fluctuations.
Prior to the 2011-April event, the only Crab $\gamma$-ray flare covered by a multi-wavelength observing program was the 2010-September flare, which triggered observations in radio, optical (using both ground-based telescopes and HST), and X-ray bands.
Despite the $\gamma$-ray brightness of the flares, there has been no evidence for correlated variations in radio (Lobanov, Horns, \& Muxlow 2011; this paper), near infrared (Kanbach, et al.\ 2010, this paper), optical (Caraveo et al.\ 2010), or X-ray bands (Evangelista et al.\ 2010; Shaposhnikov et al.\ 2010; Tennant et al.\ 2010; Ferrigno et al.\ 2010; Horns et al.\ 2010; Cusumano et al.\ 2011; Tennant et al.\ 2011; Tavani et al.\ 2011; Striani et al.\ 2011b; this paper).

Here we focus on the \Fermi-LAT results for the 2011-April flare (Buehler et al.\ 2012), which allow us to assess the source behavior in detail.
The source doubled its $\gamma$-ray flux within eight hours and reached a peak   flux 30-times its  average.
The isotropic luminosity increased to $\approx2\times10^{37}$~{\rm erg/s} in $\approx10$~hr and the spectrum peaked at $\approx400$~\MeV.
Table~\ref{t:r1} gives the $\gamma$-ray powerlaw photon spectral index, the integrated photon flux above 100~\MeV, and the photon spectral flux at 100~\MeV, as measured during the 10-ks time intervals when X-ray data (\S\ref{s:xray_obs}) were taken. 

Notification as to the level of flaring prompted us to trigger pre-approved Target of Opportunity observations with \cha~ and with the NRAO\footnote{The National Radio Astronomy Observatory is a facility of the National Science Foundation operated under cooperative agreement by Associated Universities, Inc.} Karl G.~Jansky Very Large Array (VLA).
We were also fortunate to obtain a Keck image in the near infrared, albeit not under ideal conditions.
Figure~\ref{f:r0} shows the \Fermi-LAT $\gamma$-ray counting rate as a function of time and also indicates the times of the \cha, Keck (\S\ref{s:Keck}) and some of the VLA (\S\ref{s:vla}) observations.

\section{X-ray Observations and Data Analysis} \label{s:xray_obs}

With the back-illuminated ACIS S3 CCD on the \cha~ X-ray Observatory approximately centered on the Crab pulsar, we obtained five observations (Table~\ref{t:r1}) during and somewhat after the 2011-April $\gamma$-ray flare.
For these observations, the spacecraft dithered with an amplitude set to $1\arcsec$.
Although standard processing typically produces an aspect solution better than $0.5\arcsec$, even this small uncertainty can introduce noticeable shifts when comparing different data sets.  
Thus, we re-registered images for our analysis using the read-out streak and the pulsar as guides.
As each of these 5 images was placed at approximately the same CCD location, spatial non-uniformity in the ACIS response (e.g., due to contamination) does not introduce spurious temporal variability.

Owing to the Crab's high flux, the ACIS observations employed a special mode with 0.2-s frame time, which limits the CCD read-out to a $300\times 300$~ACIS-pixel ($\approx 150\arcsec\times 150\arcsec$) subarray.
Although each observation lasted about 10 ks, telemetry saturation reduced the effective integration time to approximately 1200 s per observation.
Despite the short frame time of the special ACIS mode, regions of high surface brightness suffer somewhat from pile-up effects. 
We consider only data in the range 0.5--8.0~\keV\ because of severe interstellar absorption at low energies and declining flux at high energies. 
Using these data, we then search for X-ray variations approximately contemporaneous with the 2011-April $\gamma$-ray flare.

\subsection{X-ray Image Analysis \label{ss:x_analysis}}

Figure~\ref{f:r1} shows an image of the number of counts per ACIS pixel, summed over the 5 observations.
For each observation, we re-binned a $120\times 120$~ACIS-pixel image centered on the pulsar into a $60\times 60$~array of $2\times 2$~ACIS pixels. 
Each of these $I = 3600$ ``analysis pixels" is sufficiently large (about 1 square arcsec) to enclose most of the \cha~ point spread function anywhere in the field of view.  
Note that we also performed an analysis using a circular bin of radius $1$~ACIS pixel on an oversampled grid of cadence $0.1\times$~ACIS pixel --- i.e., a spherical top hat smoothing of the events. 
As each method gave similar results, we here report the results for the ``analysis pixel" binning, for which each pixel is statistically independent.

For each analysis pixel $i$, we calculate the mean count rate $r_i$ averaged over the $J = 5$ observations, weighted\footnote{\ $r_i = \sum_{j=1}^{J} \{r_{ij}/\sigma_{ij}^{2}\} / \sum_{j=1}^{J} \{1/\sigma_{ij}^{2}\}$} by the respective (counting-rate) statistical error $\sigma_{ij}$ for each analysis pixel and observation.
For evaluating statistical significance of temporal variations over the $J = 5$ observations, we compute\footnote{\ $\chi_{i}^{2} = \sum_{j=1}^{J} \{(r_{ij}-r_{i})^{2}/\sigma_{ij}^{2}\}$ and $S_{i} \equiv (\chi_{i}^{2}-\nu_{i})/\sqrt{2 \nu_{i}}$ where $\nu_{i} = (J-1)$.} $\chi_{i}^{2}$ along with a derived significance measure $S_{i}$.

For purposes of discussion, we also compute the appropriately weighted\footnote{\ $\sigma_{i}^{2} = J / \sum_{j=1}^{J} \{1/\sigma_{ij}^{2}\}$ and $s_{i}^2 = \frac{J}{(J-1)} \sum_{j=1}^{J} \{(r_{ij}-r_{i})^{2}/\sigma_{ij}^{2}\} / \sum_{j=1}^{J} \{1/\sigma_{ij}^{2}\} = \frac{\chi_{i}^{2}}{(J-1)} \sigma_{i}^{2}$} statistical error $\sigma_{i}$ and sample standard deviation $s_{i}$ for each pixel $i$.
As properly weighted, $\chi_{i}^{2} = (J-1)\, s_{i}^{2} / \sigma_{i}^{2}$.
While we have rigorously calculated $r_{i}$, $\chi_{i}^{2}$, $\sigma_{i}$, and $s_{i}$ for each pixel $i$ using appropriate weightings, we note that the weightings are nearly uniform as the effective duration of the each of the $J=5$ observations was nearly the same---about 1200 s.

\subsection{Variability of the X-ray images} \label{ss:x_variability}

The (counting) statistical error $\sigma_{i}$ is the primary noise term and thus governs the sensitivity for detecting temporal variations at the analysis-pixel (square-arcsec) scale.
Figure~\ref{f:r2} shows the image and the corresponding histogram of the distribution of $\sigma_{i}$, which ranges from 0.0025 to 0.024 ct/s per analysis pixel.
Based upon the $\chi^{2}$ probability distribution and the number of ``tries'' ($I=3600$ independent analysis pixels), a 99\%-confidence detection would require a $\chi_{i, 99\%}^{2} > 31.2$ on $(J-1) = 4$~degrees of freedom.
This corresponds to a sample standard deviation $s_{i, 99\%} > 2.80\, \sigma_{i}$, which ranges from 0.0071 to 0.068 ct/s over the field.

We do not here display the analogous image and histogram for the sample standard deviation $s_{i}$, which ranges from 0.0014 to 0.048 ct/s per analysis pixel.
Instead, Figure~\ref{f:r3} shows the image and histogram of the distribution of a calculated (\S \ref{ss:x_analysis}) significance measure $S_{i}$, related to $\chi_{i}^{2} = (J-1)\, s_{i}^{2} / \sigma_{i}^{2}$.
The statistically most significant variation has $\chi_{i}^{2} = 23.5$ on $\nu = (J-1) = 4$ degrees of freedom giving $S_{i} = 6.9$.
Such a fluctuation is expected statistically in at least 1 of 3600 pixels in 31\% of realizations.
Table~\ref{t:r2} gives the sample standard deviation and 99\%-confidence upper limit to the count-rate variation, for the analysis pixel with the statistically most significant X-ray variation.
While we detect no variations statistically significant at 99\%confidence, it is curious that the 3 most significant variations occur at locations on the inner ring.

Note that if a feature, such as one of the knots, doesn't change in intensity but moves from one analysis pixel to another, then our $\chi_{i}^{2}$ test would detect this as a variation.
We know that features in the inner ring of the Nebula do move and expect to detect some variability due to this motion.
However, as our 5 observations span only 14 days and $1\arcsec$ corresponds to 11.5 light days, only relativistic motion would be detectable.   
Other effects, such as changes in the roll angle of the read-out streak, can also lead to spurious variability. 
Indeed, this may play a role for the analysis pixel with the most significant variation, which lies adjacent to the average read-out streak (Figure~\ref{f:r1}).

\subsection{Limits to the X-ray flux} \label{ss:x_limits}

Thus far, we have described the X-ray data for each analysis pixel in units of ACIS count rate.
Neglecting for the moment pile-up effects, the photon spectral flux (or other related radiation quantity) is proportional to the count rate for an assumed spectral shape.
Consequently, any change in count rate corresponds to a proportionate change in photon spectral flux (for an assumed spectral shape).
Using the \cha~ PIMMS\footnote{http://asc.harvard.edu/toolkit/pimms.jsp} for the ACIS-S detector and an absorption column $N_{\rm H} = 3.1\times 10^{21}\ {\rm cm}^{-2}$, 
we determine (ignoring pile-up) this constant of proportionality for an X-ray power-law photon index $\Gamma_{x} = \frac{2}{3}$, 1, and 2:
At $E_{x} = 1$~\keV, $N_{E}(E_{x})/r =$\ 0.99, 1.26, and 2.46 $\times 10^{-3}$~ph/(cm$^2$ s \keV) per ct/s, respectively.
Correcting for pile-up has little effect in low-count-rate regions, but would raise these flux upper limits by $\approx10$\% or so for high-count-rate regions.

Table~\ref{t:r2} calculates the photon spectral flux $N_{E}(E_{x})$, the energy spectral flux $F_E(E_{x})$, and the indicative (isotropic) luminosity $E L_{E}(E_{x}) = 4 \pi D^{2} E F_{E}(E_{x})$ at $D = 2$~kpc, corresponding to the sample standard deviation and 99\%-confidence upper limit for the count-rate variation in the analysis pixel with the most significant X-ray variation.
Figure~\ref{f:r5} displays an image of the energy spectral flux $F_{E}(E_{x})$ at $E_{x} = 1$~\keV\ for $\Gamma_{x} = 1$, based upon the sample standard deviation $s_{i}$ of the count rate in each analysis pixel.

\subsection{Constraints on the X-ray to $\gamma$-ray Spectral Index} \label{ss:x_index}

We now compare the X-ray data with the $\gamma$-ray data to quantify the implications of our lack of detection of time variations in the X-ray data. 
Our approach compares a variability measure for the X-ray (1-\keV) photon spectral flux $\Delta N_{E}(E_{x})$ in each analysis pixel with the analogous variability measure for the $\gamma$-ray (100-\MeV) photon spectral flux $\Delta N_{E}(E_{\gamma})$.
In particular, we calculate the sample standard deviation of the $\gamma$-ray spectral flux at 100~\MeV, using power-law fits to the 5 \Fermi-LAT measurements that were simultaneous with the 5 \cha~ observations.
Table~\ref{t:r1} lists the 5 \cha~ ObsIDs, their dates, along with the $\gamma$-ray photon index $\Gamma_{\gamma}$, integrated photon flux $N(>$100~\MeV), and photon spectral flux $N_{E}$(100~\MeV).
For the 5 \Fermi-LAT observations, the mean and sample standard deviation of the photon spectral flux at 100~\MeV\ are $1.21\times 10^{-10}$ and $5.77\times 10^{-11}$ ph/(cm$^2$ s \keV), respectively.

Based upon the sample standard deviation ($s_{i}$) of photon spectral flux at $E_{x} = 1$~\keV\ for each X-ray analysis pixel and the measured standard deviation ($5.77\times 10^{-11}$ ph/(cm$^2$ s \keV)) at $E_{\gamma} = 100$~\MeV, we constrain the effective X-ray to $\gamma$-ray photon index of the flaring component: $\Gamma_{x\gamma} \equiv -\log[\Delta N_{E}(E_{\gamma})/\Delta N_{E}(E_{x})] / \log[E_{\gamma}/E_{x}]$.
Figure~\ref{f:r6} shows the image and corresponding histogram of the distribution of upper limits to $\Gamma_{x\gamma}$ based upon the sample standard deviation of the X-ray measurements and assuming $\Gamma_{x} = 1$.

In that the $\gamma$-ray variations are statistically significant and the X-ray variations are not, we compute 99\%-confidence upper limits to $\Gamma_{x\gamma}$ (Table~\ref{t:r2} last row).
Note that the upper limits to $\Gamma_{x\gamma}$ are marginally consistent with the low-energy extrapolation of the $\gamma$-ray spectrum ($\Gamma_{\gamma} = 1.27 \pm 0.12$) of the flaring component (Buehler et al.\ 2012).

\subsection{Variability within an X-ray Image} \label{ss:x_short}

In sections~\ref{ss:x_analysis}--\ref{ss:x_variability}, the search for variability focused on sensitivity to flux changes amongst the five pointings with a minimum cadence of 0.6~days.
Here, we search for variability on shorter time-scales---namely within each pointing. 
As described in Section~\ref{s:xray_obs}, the ACIS S3 CCD was read at most roughly 6000 times in each pointing due to telemetry saturation (deadtime). 
In this study, rather than using the $2\times2$ ACIS analysis pixel, we 
employed a circular search bin with a radius of $1$~ACIS-pixel
($0.49\arcsec$) on a grid with $0.1$~ACIS pixel spacing.
Note that this oversampling implies that the results of the test in adjacent 
pixels are not statistically independent.
(We also analyzed these data using the statistically independent analysis 
pixels of \S\ref{ss:x_analysis} with similar results as below.) 

Using the frame number of each detected photon, we derive the empirical cumulative distribution function (ECDF) of the frames with a photon arriving in the analysis pixels of the CCD. 
This ECDF is then compared with the corresponding ECDF of the exposure given 
by the sequence of frames actually read. 
Finally, we compare the two ECDFs using a Kolmogorov-Smirnov test, resulting
in a probability estimate $Q$ that the two ECDFs are derived from the same 
parent distribution. 
A low value of $Q$  would indicate possible variability. 
The results of the test for the five pointings were very similar.
The smallest value, $Q_\mathrm{min}= 6.7\times10^{-7}$, was obtained in  observation $13151$.
Note that selecting this point represents a tuning bias, as the noise in neighboring points is highly correlated due to the oversampling. 
The probability of finding at least one pixel with $Q_\mathrm{min}$ 
considering that there are $120\times120/(\pi\times1^2)\times5$ statistically independent trials is 0.015, which we regard as a {\em lower} limit due to the tuning bias. 
A 0.015 probability is tantalizing but not compellingly significant:
Hence, we do {\em not} claim detection of short-time-scale variability.
The fact that the location of the point with minimum Q is very close to the pulsar, a region in which pileup plays a strong role in blotting out the image, bolsters our somewhat conservative conclusion. 

\section{Near-Infrared Image of the Inner Knot}\label{s:Keck}

The extreme saturation of the pulsar in the X-ray images means that we cannot easily study the central 2$^{\prime\prime}$ in X-rays.
However, this region does contain a nebular structure of particular interest: the ``inner knot'' whose peak is $0.65^{\prime\prime}$ southeast of the pulsar at position angle $118^\circ$ East from North (Hester 2008). 
This structure, an oval shape extending $\approx 0.75^{\prime\prime}$, is well measured in HST and ground-based near-IR images.
Given its relatively red spectrum (energy spectral index $\alpha_\nu = -1.3 \pm 0.1$ versus $\alpha_\nu = 0.27 \pm 0.03$ for the pulsar; Sandberg and Sollerman 2009), it is one of the near-IR brightest structures in the Nebula. 
Sandberg and Sollerman (2009) note that the knot varies by a factor of 2; we confirm typical variability of $20-30\%$ in archival {\it HST} images.  
Komissarov \& Lyutikov (2011) have proposed that this structure represents radiation from an oblique termination shock in the pulsar wind nebula. 
In this picture, the Earth line-of-sight is tangent to the flow at the inner knot position, and thus the intensity experiences substantial Doppler boosting for synchrotron emission in the mildly relativistic post-shock flow. 
Indeed, in relativistic MHD simulations they find that this bright spot is highly variable and can dominate the $\gamma$-ray synchrotron emission.
Alternatively, the knot could be a time varying standing shock in the polar jet flow itself, a flow known to be highly variable from HST imaging (Hester 1995, 2002, 2008). 

It is thus of interest to check the status of the knot during the
2011-April $\gamma$-ray flare. 
Unlike the sequence of multiwavelength observations performed after the 2010 September flare, it was impossible to trigger an allocated HST Target of Opportunity observation owing to solar constraints in April.
Happily we were able to obtain a Keck Near Infrared Camera (NIRC2) $K'$ exposure (Figure~\ref{f:k1} left image) on MJD 55667.250, almost precisely at the peak of the $\gamma$-ray flux and 2.5\,h before the ACIS image ObsID 13152 (Figure~\ref{f:r0}).
Unhappily, the observations occurred during twilight and only one $20\times4$-s integration without dithering was obtained.
Under these conditions the adaptic-optics (AO) loop did not close, leaving an undithered image with native 0.46$^{\prime\prime}$ full width at half maximum (FWHM) seeing. 
This frame was dark subtracted and an approximate background was removed using an immediately subsequent image.
Despite the modest image quality, the inner knot was well detected. 
After subtracting the pulsar with a scaled image of the comparably bright companion star $4^{\prime\prime}$ northeast, we measured the knot flux and position. 
We find a magnitude $K'= 15.60\pm 0.03$ and an offset $0.64\pm0.04^{\prime\prime}$ from the pulsar. 
For comparison we measured a high-quality NIRC2 $K'$ image (Figure~\ref{f:k1} right image) obtained 2005 Nov 10. 
Here the knot is $K'= 15.94\pm 0.02$ at offset $0.58\pm0.02^{\prime\prime}$.
We also note that Sandberg and Sollerman (2009) measured $K_s = 15.80\pm0.03$ on 2003 Oct 18. 
We conclude that the knot was in a relatively bright state during the flare ($\approx 35\%$ brighter than in 2005), but well within the normal range of flux (and position) variation. 
Thus, there is no dramatic change in the inner knot in the near-IR band.
We use the amplitude of the measured variation as an upper limit to any variation in the inner knot associated with the $\gamma$-ray flare (Figure~\ref{f:sed}).

\section{Radio Observations}\label{s:vla}

On 2011 April 14, we triggered a prompt radio follow-up program with the VLA. 
The VLA observations occurred in 8 epochs starting April 15 and ending July 10.
These observations detected the pulsar at 2 epochs, but found no other point sources in the field.  
Observations were predominantly in the range $4-8$~GHz, with additional observations at 1.4~GHz  (not reported; see below) and at 22~GHz (for some later epochs). 
Unless otherwise noted, all observations used two sub-bands, each with a 128-MHz bandwidth.
In each run, observations of the target were bracketed with scans of a phase calibrator (J0559+2353, except where noted) and a flux calibrator (3C~147). 
The fields of view are limited to the primary beam response of the antennas, with full width at half power of 9$'$/$\nu_{\rm 5}$ with $\nu_{\rm 5} \equiv \nu$/(5~GHz). 
Table~\ref{t:vla1} provides a summary of the observational parameters and results.

Our initial observations were obtained through a \Fermi~ guest-investigator cycle-3 program (S3184) approved for four 1-hour runs in the L-band and C-band ($\approx\, 1.4$ and $5$ GHz, respectively). 
The VLA was in its B-array configuration during these observations, resulting in images with angular resolution $\approx\,$1\arcsec/$\nu_{5}$. 
We found that the L-band data for the Crab were highly confused due to the brightness and complexity of the steep-spectrum nebular emission in the first (April 15) and second (April 19) epochs.
Consequently, we modified our strategy for subsequent observations. 
After the first epoch, we split the C-band observations into two widely spaced side-bands centered at 4.2 and 7.8 GHz, aiming better to constrain the spectrum of any detected source. 
We also began scheduling observations only at frequencies greater than 4~GHz after the second epoch. 
In these B-array data, our point-source limits at the lower frequency are $\approx 3-4\times$ larger than at the higher frequency, again due to the Crab Nebula's steep-spectrum radio emission.

After non-detection of any significant radio point-source emission down to $\approx\,$1-7 mJy (3-sigma) sensitivities in the initial 
three VLA observations 1--7 days after the $\gamma$-ray peak (Hays et al.\ 2011), we purposely delayed the fourth observation until 9 days after the previous observation to probe longer time scales.  
Only in this last observation (April 30) did we obtain a significant point-source detection, which was coincident with the Crab pulsar position, but only in the upper side-band centered at 7.8~GHz. 
The detection is a factor of 5 greater than the 3-sigma limit from 9 days prior. The source was not detected at the lower frequency side-band (4.2~GHz) with a limit indicating a source with a flat radio spectrum.

Following the point-source detection on April 30, we became aware of VLA TEST observations of the Crab obtained  on April 22 (program TDEM0007, PI: D.~Frail). 
These data were obtained with wide bandwidth (16 $\times$ 128 MHz wide sidebands), so were more sensitive than those from our observing runs. 
The flux densities were scaled to 3C~147; J0534+1927 was utilized for phase calibration. 
The flat-spectrum radio source coincident with the Crab pulsar detected in our Apr-30 observation was confirmed in the Apr-22 data in two bands, but with a much lower ($10\times$) flux. 
Also, the source spectrum was rather steep, with an energy spectral index $\alpha = 2.21\pm 0.34$ ($S_{\nu}\propto\nu^{-\alpha}$) between 5~GHz and 8.6~GHz.

Following the radio detections of the pulsar, we requested further VLA monitoring of the Crab through Director's Discretionary Time (program 11A-268 = AC1052). 
In addition to the C-band observations, we obtained exposures in the K~band (centered at 22.396 and 22.254 GHz) aiming to constrain further the spectrum of any detected radio source. 
Through this program, we obtained 2-hour runs on May 12/13 (while the VLA was in its hybrid BnA array) and on July 10/11 (in A array), and an additional 1-hour run on May 28, using one of the early (April 19, from program S3184) frequency setups. 
An angular resolution of $\approx$(0.3\arcsec/$\nu_{\rm 5}$) is typically achieved in A-array VLA observations. 
With the higher resolution, we obtained systematically 4$\times$ lower flux limits than in the lower resolution B-array data, presumably due to lesser contribution from the extended nebular emission. 
In none of these later epoch follow-up observations, did we detect a point source, to typical limits of $1-2$ mJy at each of the three frequencies (see Table~\ref{t:vla1}).

\subsection{Discussion of the Radio Data} \label{ss:vladiscussion}

Previously, it was argued that the $\gamma$-ray flaring possibly originates in a knot 5.7\arcsec\ east of the pulsar (Tavani et al.\ 2011). 
Indeed, this knot is the site of the most significant X-ray variability we observe (\S~\ref{ss:x_variability}) during the 2011-April flaring episode. 
Variable radio emission was detected around the time of the previous Crab $\gamma$-ray 
flaring episode (Lobanov, Horns, \& Muxlow 2011) with fainter flux densities than achieved in our VLA observations. 
However, we found no significant radio point-source counterpart to this knot in any of our 8 epoch VLA observations following the 2011-April $\gamma$-ray flare. 
Rather, we detected a variable continuum radio source with the VLA, coincident within $0.2\arcsec$ of the Crab pulsar position in 2 of 8 epochs. 
However, the flux level and cadence of the radio detections is consistent with previous observations by Moffett \& Hankins (1996), who detected the pulsar $ 20-40\%$ of the times they observed. 
Moreover, the dates of the radio point-source detections coincident with the pulsar do not coincide with any feature in the $\gamma$-ray lightcurve (Figure~\ref{f:r0}), having occurred 8-16 days after the brightest $\gamma$-ray peak. 
Consequently, our VLA follow-up observations provide no conclusive evidence for the site of the $\gamma$-ray flares.

\section{Discussion} \label{s:consequences}

Here we discuss possible explanations for the absence in non-$\gamma$-ray bands of variability that is obviously correlated with the $\gamma$-ray flare. In addition we present a conceptual model for the production of the $\gamma$-ray flares.

\subsection{$\gamma$-ray Emission}

As was recognized immediately, the SED of $\gamma$-ray flares peak near a characteristic energy about 5 times the energy $\alpha^{-1}m_ec^2\approx 70$~\MeV, which is identified with radiation-reaction-limited synchrotron (magneto-bremsstrahlung) emission (e.g., Landau \& Lifshitz 1959). 
Subjected to comparable parallel and perpendicular electromagnetic acceleration, an electron radiates at this energy, independent of the strength of the acceleration. 
An electron emitting synchrotron radiation in a magnetostatic field with peak emission at several $100$~\MeV\ would cool in turning through $\approx0.2$~radian and would thus require a parallel electric field $E\approx5 cB$ to compensate the radiative loss.

The Crab pulsar releases energy in an essentially electromagnetic form. 
Poynting flux flows radially outward from the the pulsar through the light cylinder at    
$c P/(2\pi) \approx 1500$~km and into an outflowing wind, where at least some of the electromagnetic-energy flux may transform into a plasma-energy flux. 
How, where, and to what extent this happens has long been a matter of debate (e.g., Arons 2010; Kirk et al.\ 2009).
Furthermore, the electromagnetic component has a DC toroidal part with an associated quadrupolar current distribution, and an AC, ``striped'' part containing current sheets separated by $\frac{1}{2} c P \approx5000$~km. 
The transformation from electromagnetic to plasma energy might be non-dissipative---through the action of a Lorentz force (e.g., Bogovalov 1997; Bogovalov 2001)---or dissipative---through particle heating and acceleration (e.g., Coroniti 1990; Lyubarsky \& Kirk 2001; Sironi \& Spitkovsky 2011).  
However, it must occur somewhere as magnetic flux would otherwise accumulate in the nebula, ultimately reacting back on the pulsar.
Some of this transformation from electromagnetic to plasma energy may occur at a shock (P{\'e}tri \& Lyubarsky 2007;  Sironi \& Spitkovsky 2011) with radius $\approx10^{17}$~cm, where the wind momentum flux balances the ambient nebular pressure (Rees \& Gunn 1974; Kennel \& Coroniti 1984). 
It has also been proposed that the toroidal field loops contract to form an axial pinch (identified with the X-ray jet) and reconnect at an equatorial current sheet (the torus) (Komissarov \& Lyubarsky 2003; Del Zanna, Amato, \& Bucciantini 2004; Camus et al.\ 2009).

In many respects the pulsar is a current generator. 
The supersonic wind contains outflowing fluxes of electrons and positrons. 
(Any ions that are present behave like positrons of similar rigidity but do not radiate.)
The difference in their fluxes determine the current density. 
This current may concentrate into sheets and filaments, where strong dissipation can occur---as happens in heliospheric and laboratory plasmas (Gosling et al.\ 2005; Sui \& Holman 2003; Sergeev et al.\ 1993).
In particular, the inner  wind, the shock, the jet, and the torus are all natural sites of rapid dissipation and $\gamma$-ray emission. 
If we consider this dissipation more generally under electromagnetic conditions, a current $I$ may be associated with a potential difference $V\approx IZ_0$ where $Z_0=\mu_0 c=377\, \Omega$ is the impedance of free space and we drop model-dependent constants of order unity. 
(This result can be anticipated on the basis of dimensional analysis or exhibited in particular simple cases.) 
The maximum energy to which an electron or positron can be accelerated is $\gamma_{{\rm max} }m_e c^2\approx$ \eV\ $\approx e I Z_0$ and the expected power is then $L\approx I V\approx I^2 Z_0$. 
On this basis, a spectrum of currents extending up to $\approx30$~TA should suffice to account for the $\gamma$-ray variations. 

However, currents do not automatically dissipate as just described. 
Large electric fields are normally discharged in a few plasma periods.
The best way to create them here is transiently over a few Larmor periods and radii. 
This, in turn, requires local charge separation of the plasma in the emission site.
To be more precise, the density $n=n_-+n_+$ of electrons plus positrons/ions will combine to create a local current density $j\approx nec$. 
For example, in the case of a pinch, the gradient, polarization and curvature drifts automatically produce the axial current.  
However, a supporting local electric field of strength $E\approx cB$ also requires that $|n_--n_+|\approx n$. 

Put another way, the charge and current are mostly in an emission site that is a few Larmor radii in size and survives for a few Larmor periods of the $\gamma$-ray emitting particles. 
When this happens, the ``Ohmic'' dissipation is radiative, not collisional as is normally the case. 
For this to occur, the particles must be sufficiently energetic to radiate efficiently. 
This requires that most of the particles are concentrated in an emission site that is as small as $\approx\gamma_{{\rm max}}^3r_e\approx(eIZ_0 /m_ec^2)^3r_e\approx10^{16}$~cm.  
The key point is that there should be extensive and sustained radiation-reaction-limited emission at the peak $\gamma$-ray-flare energy, even though the particle energy and magnetic field might be changing.  
Note that when this condition is unsatisfied, efficient particle acceleration to lower energy should still result: Most of the magnetic dissipation and particle acceleration in the nebula might occur in this fashion.
Detailed modeling is necessary to determine whether or not such a scheme can reproduce the powerful, narrow-band $\gamma$-ray variation that is observed (Uzdensky et al.\ 2011; Cerutti et al.\ 2012; Cerutti, Uzdensky, \& Begelman 2012; Lyutikov, Balsara, \& Matthews 2012; Bykov et al.\ 2012; Sturrock \& Aschwanden 2012; Blandford \& Yuan 2012) and to see if unstable magnetized plasmas, carrying large currents, evolve to satisfy these conditions. 

\subsection{Associated Emission}

Whether we interpret the $\gamma$-rays as coming from radiation-reaction-limited synchrotron emission or simply extrapolate the observed $\gamma$-ray spectra to lower energy, it should not be surprising that direct, associated emission has not yet been observed in the X-ray, optical, or radio bands:  
In these bands the contrast with the steady emission is too small to be easily noticed. 
However, the indirect effects could be larger and detectable. 
For example, the large 2011-April flare produced a radiant energy of $6\times10^{40}$~ergs if isotropic, equivalent to the energy contained within a region of size $\approx2\times10^{16}$~cm subtending an angle $\approx0.3$~arcsec. 
It seems unlikely that the dynamical aftermath of a major flare would not alter the ambient emission---either through  compression or rarefaction that would cause the magnetic field strength 
and the electron distribution function to change significantly. 
The associated surface brightness change should be several percent, assuming a total emission region of size $\approx0.3$~arcsec, consistent with our upper limits.
Even if future observations fail to exhibit associated emission, they may still rule out specific detailed mechanisms in local sites. 

Understanding the emission mechanism could have a significance beyond pulsar wind nebulae. 
In particular, it could provide a clue to the surprisingly rapidly variable emission seen in relativistic jets in the radio, optical, X-ray, and \TeV\ bands. 
If so, the Crab Nebula would once again be the source of fresh and important astrophysical insight.  

\section{Summary}\label{s:summary}

Using the \cha, Keck, and VLA Observatories, we acquired X-ray, near-IR  and radio images of the Crab Nebula, contemporaneous with the 2011-April $\gamma$-ray flare. 
We searched for variability in the X-ray data over two time-scale ranges: 
First we tested for pointing-to-pointing variations amongst the 5 pointings, each with an effective exposure time $\approx 1200$~s and a minimum separation of 0.6~days. 
Second we tested for variations within each of the 5 observations.
In neither case did we detect statistically significant X-ray variations; thus we can set only upper limits to any X-ray variations associated with the $\gamma$-ray flare.
As the \cha~ ACIS images suffer severe pile-up near the Crab pulsar, our search for variability in the X-ray images was not sensitive to variations within the central $\approx 1.5\arcsec$ or so.

Comparing the upper limits to X-ray variations with the \Fermi-LAT-measured $\gamma$-ray variations, we set upper limits at $99$\%-confidence to the effective X-ray--$\gamma$-ray photon power-law index $\Gamma_{x\gamma} \leq 1.20$ to $ \leq 1.27$, dependent upon assumptions about the X-ray index $\Gamma_{x}$.
As \Fermi-LAT measures a $\gamma$-ray index $\Gamma_{\gamma} = 1.27\pm 0.12$ for the flaring component, it is statistically possible that the flaring component's spectrum extends as a simple power-law from $\gamma$-rays to X-rays.
Further, we note that our upper limit to $\Gamma_{x\gamma}$ is consistent with transparent synchrotron emission, whose photon index must be $>\frac{2}{3}$.

Comparison of two Keck near-IR observations found that the inner knot ($\approx 0.65\arcsec$ from the pulsar) was somewhat brighter than average during the $\gamma$-ray flare, but well within the normal range of brightness fluctuations typically observed.
We used the measured ($\approx35\%$) change in the near-IR flux from this knot as an upper limit to near-IR variations associated with the $\gamma$-ray flare.
We also performed a number of VLA observations searching for a point source appearing either at an unusual location and/or contemporaneous with the $\gamma$-ray flare. 
Other than the pulsar itself, no such source was detected.

Figure~\ref{f:sed} shows the spectral energy distribution (SED) of the Crab Nebula over the observed electromagnetic spectrum.
The plot also shows the SED of the 2011-April $\gamma$-ray flare and the various limits determined here on variable radio, near-infrared, and X-ray emission possibly associated with the $\gamma$-ray flare.

Finally, we reviewed and discussed potential implications of $\gamma$-ray flares and theoretical issues to be addressed.
We concluded that, apart from lower-energy emission directly associated with the $\gamma$-ray flare itself, the dynamical aftermath of a major flare could alter the ambient emission --- e.g., through compression of the magnetic field. 
The associated surface brightness change would likely be only several percent, assuming a total emission region of size $\approx0.3\arcsec$, consistent with our upper limits.

Although no ``smoking gun" has been identified, one should be encouraged that we have identified a number of regions in the X-ray images that are possible candidates. 
We have also established further Target of Opportunity observations with \cha~ and HST that will be triggered at the onset of the next $\gamma$-ray flare. 
The X-ray observations will also probe the region very close to the pulsar using the \cha~ High-Resolution Camera (HRC). 

\section{Acknowledgments}

The work of MCW, SLO, and AFT is supported by the \cha~ Program.
The \cha~ data was obtained in response to a pre-approved target of opportunity request granted under \cha~ Director's Discretionary Time.
The \Fermi~ LAT Collaboration acknowledges generous ongoing support
from a number of agencies and institutes that have supported both the
development and the operation of the LAT as well as scientific data analysis.
These include the National Aeronautics and Space Administration and the
Department of Energy in the United States, the Commissariat \`a l'Energie Atomique
and the Centre National de la Recherche Scientifique / Institut National de Physique
Nucl\'eaire et de Physique des Particules in France, the Agenzia Spaziale Italiana
and the Istituto Nazionale di Fisica Nucleare in Italy, the Ministry of Education,
Culture, Sports, Science and Technology (MEXT), High Energy Accelerator Research
Organization (KEK) and Japan Aerospace Exploration Agency (JAXA) in Japan, and
the K.~A.~Wallenberg Foundation, the Swedish Research Council and the
Swedish National Space Board in Sweden.
Additional support for science analysis during the operations phase is gratefully
acknowledged from the Istituto Nazionale di Astrofisica in Italy and the Centre National d'\'Etudes Spatiales in France.
CCC, GBT, and JBL thank Tim Hankins for useful discussions, and the NRAO scheduling 
committee for alerting us to the VLA TEST data and for their prompt consideration of our Director's Discretionary Time observations.
CCC, GBT, and JBL were supported in part by NASA through a \Fermi~ cycle-3 guest investigator grant. 
In addition, work by CCC~at NRL is supported in part by NASA DPR S-15633-Y. 
Our analyses utilized software tools provided by the \cha~ X-ray Center (CXC) in the application package CIAO and from the High-Energy Astrophysics Science Archive Research Center (HEASARC, operated by the NASA Goddard Space Flight Center, Greenbelt MD, and by the Smithsonian Astrophysical Observatory, Cambridge MA).

{\it Facilities:} \facility{CXO}, \facility{Fermi}, \facility{VLA}

\clearpage

\begin{table}
\begin{center}
{\caption {Time-ordered list of \cha\ observations and \Fermi-LAT $\gamma$-ray measurements$^a$. \label{t:r1}}}
\begin{tabular}{lc|ccr} \\ \hline 
 ObsID & Date $^b$  & $\Gamma_\gamma ^{\ c}$   & $N(>$ 100 \MeV)$^{\ d}$           & $N_{E}$(100 \MeV)$^{\ e}$ \\ \hline
 13150 & 4851.039   & $2.42\pm 0.08$           & $(1.27\pm 0.08)\times 10^{-5}$    & ($1.80\pm 0.15)\times 10^{-10}$ \\ 
 13151 & 4851.667   & $2.25\pm 0.15$           & $(8.05\pm 0.15)\times 10^{-6}$    & $(1.01\pm 0.12)\times 10^{-10}$ \\ 
 13152 & 4853.423   & $2.20\pm 0.06$           & $(1.54\pm 0.06)\times 10^{-5}$    & $(1.84\pm0.12)\times 10^{-10}$ \\ 
 13153 & 4859.032   & $2.27\pm 0.21$           & $(4.68\pm 0.21)\times 10^{-6}$    & $(6.0\pm 1.0)\times 10^{-11}$ \\
 13154 & 4865.335   & $2.76\pm 0.40$           & $(4.50\pm 0.40)\times 10^{-6}$    & $(7.9\pm1.9)\times 10^{-11}$ \\ \hline
\end{tabular} 
\end{center}
$^a$ Analyzed following all the procedures in Buehler et al.\ (2012) \\
$^b$ Days after MJD 50814 to the middle of the observation, which is 10-ks long. \\
$^c$ $\gamma$-ray powerlaw number index. The error is the maximum of the two-sided uncertainty\\
$^d$ Photon integrated flux [ph/(cm$^2$ s)] above 100~\MeV. The error is the maximum of the two-sided uncertainty\\
$^e$ Photon spectral flux [ph/(cm$^2$ s \keV)] at 100~\MeV \\
\end{table}

\begin{table}
\begin{center}
{\caption {X-ray results at 1~\keV\ for the analysis pixel with the most significant variation. \label{t:r2}}}
\begin{tabular}{ll|ccc|ccc|} \\ \hline
\multicolumn{1}{l}{Quantity} & \multicolumn{1}{l|}{Unit} & \multicolumn{3}{c|}{Sample stdev ($s$)} 
& \multicolumn{3}{c|}{99\%-upper limit} \\  \hline
\multicolumn{1}{l}{Rate} & \multicolumn{1}{l|}{ct/s} & \multicolumn{3}{c|}{0.0480} 
& \multicolumn{3}{c|}{0.0554} \\            
$\Gamma_{x}$ &                                & $\frac{2}{3}$ & 1  & 2   & $\frac{2}{3}$ & 1  & 2      \\ \hline
$N_{E}$      & $10^{-4}$ ph/(cm$^2$ s \keV)   & 0.48   & 0.61   & 1.18   & 0.55   & 0.70   & 1.36   \\
$F_{E}$      & $10^{-13}$ erg/(cm$^2$ s \keV) & 0.76   & 0.97   & 1.89   & 0.88   & 1.12   & 2.18   \\
$E L_{E}$    & $10^{32}$ erg/s                & 0.37   & 0.47   & 0.91   & 0.42   & 0.54   & 1.05   \\
$\Gamma_{x\gamma}$ &                          & 1.18   & 1.20   & 1.26   & 1.20   & 1.22   & 1.27   \\ \hline
\end{tabular} 
\end{center}
\end{table}

\clearpage

\begin{table}
\begin{center}
\caption{VLA point-source detections and limits in the Crab field\label{t:vla1}}
\vspace{0.1in}
\begin{tabular}{lcccc}
\hline
\hline
\multicolumn{1}{l}{Epoch} &
\multicolumn{1}{c}{Date$^{a}$} &
\multicolumn{1}{c}{Array} &
\multicolumn{1}{c}{Frequency} &
\multicolumn{1}{c}{Detection or Limit$^{c}$} \\
\multicolumn{1}{l}{(in 2011)} &
\multicolumn{1}{c}{} &
\multicolumn{1}{c}{} &
\multicolumn{1}{c}{(GHz)} &
\multicolumn{1}{c}{(mJy)} \\
\hline
Apr 15 & 4852.980 & B   & 4.959 & $<$ 7.1 \\  
Apr 19 & 4856.006 & B   & 4.195 & $<$ 5.7 \\  
Apr 19 & 4856.006 & B   & 7.795 & $<$ 1.7 \\  
Apr 21 & 4858.011 & B   & 4.195 & $<$ 4.3  \\ 
Apr 21 & 4858.011 & B   & 7.795 & $<$ 1.0  \\ 
Apr 22 & 4859.919 & B   & 4.910 & 1.78 $\pm$ 0.086\\       
Apr 22 & 4859.919 & B   & 8.566 & 0.52 $\pm$ 0.094 \\      
Apr 30 & 4867.901 & B   & 4.195 & $<$ 7.9 \\               
Apr 30 & 4867.901 & B   & 7.762$^{b}$ & 5.55 $\pm$ 0.86 \\ 
Apr 30 & 4867.901 & B   & 7.827$^{b}$ & 4.59 $\pm$ 0.96 \\ 
May 12 & 4879.985 & BnA & 4.195 & $<$2.1 \\ 
May 12 & 4879.985 & BnA & 7.795 & $<$0.8 \\ 
May 12 & 4879.985 & BnA & 22.46 & $<$0.4 \\ 
May 28 & 4895.925 & BnA & 4.195 & $<$2.4 \\ 
May 28 & 4895.925 & BnA & 7.795 & $<$1.2 \\ 
Jul 10 & 4938.791 & A   & 4.195 & $<$0.7 \\ 
Jul 10 & 4938.791 & A   & 7.795 & $<$1.1 \\ 
Jul 10 & 4938.791 & A   & 22.46 & $<$1.7 \\ 
\hline
\end{tabular}

\end{center}
Notes -- \\
$^{a}$ Days after MJD 50814 to the middle of the on-target exposures.\\
$^{b}$ For the 7.8 GHz detection on Apr 30, the data were further split into two sub-bands.\\
$^{c}$ Limits list 3-sigma uncertainties. 
Detections list 1-sigma uncertainties; corresponding
point-source limits during these 2 epochs should take 3 times the 1-sigma values.
\end{table}

\clearpage

\begin{center}
\includegraphics[angle=-90,width=17.0cm]{fig_fermi_lc_april.ps}
\figcaption{\Fermi-LAT photon flux ($10^{-7}$~ph/(cm$^2$~s)) above 100~\MeV\ during the 2011-April flare as a function of time. 
Displayed data extend beyond the time span shown in Buehler et al.\ (2012) but follow the same data processing as described there: 
Data are adaptively binned with a 20-minute average bin duration. 
The full-range vertical lines denote times of the 5 ($\approx$ 1200-s) \cha~ observations (black), of the Keck observation (blue, fourth from left), and of the first 5 VLA observations (green).
\label{f:r0}
}
\end{center}

\clearpage

\begin{center}
\includegraphics[angle=-90,width=17.0cm]{fig_image_r_w_sgt6.ps}
\figcaption{Summed image for the 5 \cha\ ACIS observations occurring near the 2011-April $\gamma$-ray flare, at the native CCD resolution.
The color bar gives summed counts per ACIS pixel over a total effective exposure of about 6 ks.
North is up and the pulsar is at (0,0) in the displayed  ACIS-pixel coordinates. 
The nearly horizontal read-out streak through the pulsar's location is the trailed (out-of-time) image, resulting from exposure of each CCD pixel as the image is read out at 40~$\mu$s per row.
As the 5 observations occurred at slightly different roll angles, the read-out streak is slightly blurred azimuthally. 
The $\mathsf X$ symbols mark locations of the 3 statistically most significant variations ($S_{i} > 6$, \S \ref{ss:x_variability}), the most significant lying to the east of the pulsar.
\label{f:r1}
}
\end{center}

\begin{center}
\includegraphics[angle=-90,width=8.0cm]{fig_er_image.ps}
\includegraphics[angle=-90,width=8.cm]{fig_noi_vs_er.ps}
\figcaption{Left: Image of the statistical error $\sigma_{i}$ (ct/s) in the counting rate per analysis pixel (about 1 square arcsec) for the 5 \cha\ ACIS observations. 
North is up and the pulsar is at (30,30) in the displayed analysis-pixel coordinates. 
Right: Histogram of number of occurrences of each value in the image to the left. 
\label{f:r2}
}
\end{center}

\begin{center}
\includegraphics[angle=-90,width=8.0cm]{fig_s_image.ps}
\includegraphics[angle=-90,width=8.0cm]{fig_noi_vs_s.ps}
\figcaption{Left: Image of the significance measure $S_{i} \equiv (\chi_{i}^{2}-\nu_{i})/\sqrt{2 \nu_{i}}$ of sample counting-rate variations amongst the 5 observations. 
North is up and the pulsar is at (30,30) in the displayed analysis-pixel coordinates.
Right: Histogram of number of occurrences of each value in the image to the left.
\label{f:r3}
}
\end{center}

\begin{center}
\includegraphics[angle=-90,width=9.0cm]{fig_f_1p0_image.ps}
\figcaption{Image of the energy spectral flux $F_{E}(E_{x})$, in erg/(cm$^2$ s \keV) at $E_{x} = 1$~\keV, based upon the sample standard deviation ($s_{i}$) of the counting rate and assuming $\Gamma_x = 1$.
Note that the indicative energy flux $F(E) \equiv E F_{E}(E)$ in erg/(cm$^2$ s) happens to have the same numerical value as $F_{E}(E)$ at $E = 1$~\keV.
North is up and the pulsar is at (30,30) in the displayed analysis-pixel coordinates.
\label{f:r5}
}
\end{center}
\clearpage

\begin{center}
\includegraphics[angle=-90,width=8.0cm]{fig_gxg_1p0_image.ps}
\includegraphics[angle=-90,width=8.0cm]{fig_gxg_1p0_noi.ps}
\figcaption{Left: Image of upper limits to the effective photon index $\Gamma_{x\gamma}$ between 1~\keV\ and 100~\MeV, based upon the sample standard deviation ($s_{i}$) of the counting rate and assuming $\Gamma_x = 1$.
North is up and the pulsar is at (30,30) in the displayed analysis-pixel coordinates.
Right: Histogram of number of occurrences of each value in the image to the left.
\label{f:r6}
}
\end{center}

\begin{center}
\includegraphics[angle=0,width=17.0cm]{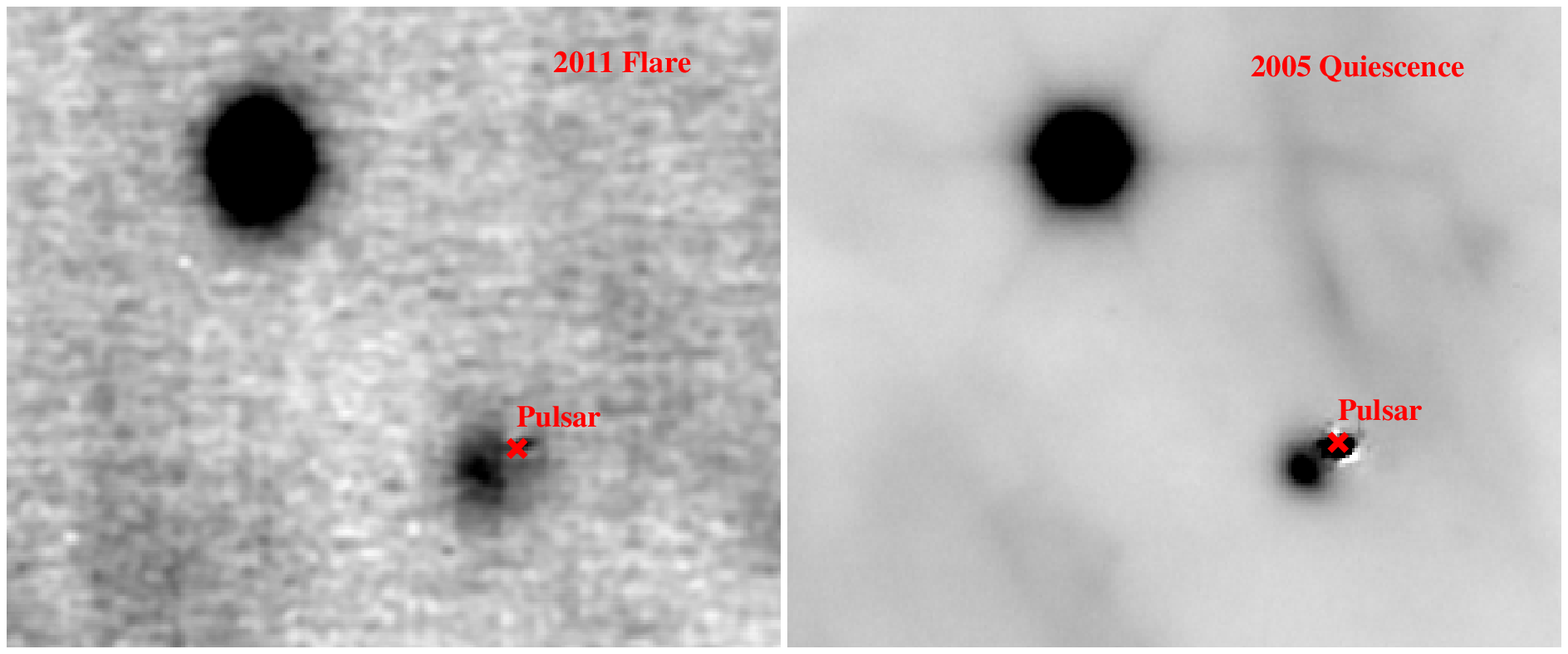}
\figcaption{Keck $K^\prime$ observations of the Crab, after subtraction of a field-star image from the pulsar position (marked with $\mathsf{x}$). 
Left: (MJD-50814) = 4853.250 without adaptive optics (AO). 
Right: (MJD-50814) = 2870, with laser guide star AO. 
Residuals from imperfect subtraction of the point spread function are visible at the pulsar position;
the `inner knot' is the extended structure to the southeast.
\label{f:k1} }
\end{center}

\begin{center}
\includegraphics[angle=270,width=17.0cm]{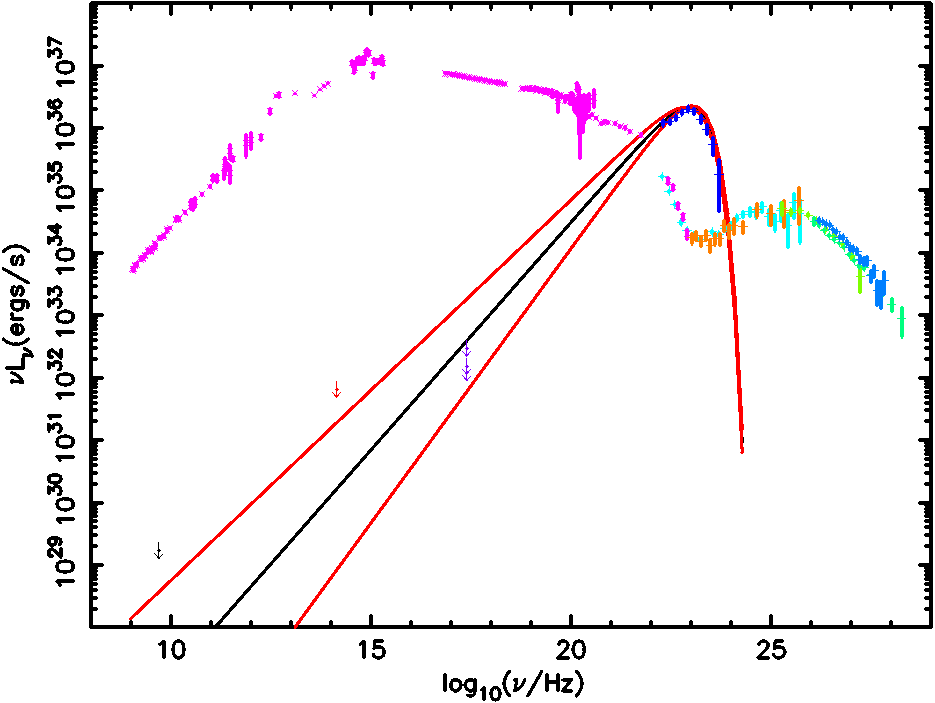}
\figcaption{Spectral energy distribution (SED) of archival data (purple, orange, cyan, light green, turquoise, light blue) compiled by Meyer, Horns, \& Zechlin (2010). 
Power is scaled from flux assuming isotropic emission at 2~kpc. 
The \Fermi-LAT data for the 2011-April flare component appear in dark blue (Buehler et al. 2012). 
The solid black and red curves are fits to the flare spectrum with a power-law extrapolation to lower energies of photon index $\Gamma = 1.27\pm 0.12$ (spectrum \textit{7} in Buehler et al. 2012).
The three downward blue arrows at $\log_{10}\nu=17.4$ mark 99\%-confidence upper limits to a variable X-ray component, in increasing $E L_{E} = \nu L_\nu$ for $\Gamma_{x}= \frac{2}{3}$, 1, and 2 respectively.
(NB: Values for $E L_{E}$ from Table~\ref{t:r2} are multiplied by $2\sqrt{2}$ to scale from a standard deviation to a peak-to-valley, for comparison with the plotted SED of the $\gamma$-ray flare.)
The red downward arrow at $\log_{10}\nu=14.1$ indicates an upper limit to infrared variability of the inner knot, determined from the difference between the two Keck images.
Finally, the black downward arrow at $\log_{10}\nu=9.7$ gives an upper limit to 5-GHz radio variability, based upon the April-15 VLA measurement. 
\label{f:sed}
}
\end{center}

\end{document}